\documentclass[12pt,preprint]{aastex} 
\usepackage{bm}
\usepackage{amsmath}
\usepackage{float}

\shorttitle{} 
\shortauthors{} 
 
\begin{document} 
 
\received{} 
\accepted{} 
 
\title{Tidally Induced Pulsations in Kepler Eclipsing Binary KIC~3230227}  
 
\author{Zhao Guo, Douglas R. Gies} 
\affil{Center for High Angular Resolution Astronomy and  
 Department of Physics and Astronomy,\\ 
 Georgia State University, P. O. Box 5060, Atlanta, GA 30302-5060, USA; \\guo@astro.gsu.edu, gies@chara.gsu.edu } 
 
\author{Jim Fuller}
\affil{TAPIR, Walter Burke Institute for Theoretical Physics, Mailcode 350-17, Caltech, Pasadena, CA 91125, USA;}
\affil{Kavli Institute for Theoretical Physics, Kohn Hall, University of California, Santa Barbara, CA 93106, USA;\\jfuller@caltech.edu}

%\author{Rachel A. Smullen}
%\affil{Steward Observatory, University of Arizona, Tucson, AZ 85721, USA;\\ rsmullen@email.arizona.edu}
\slugcomment{10/25/2016} 
%\paperid{}

%%%%%%%%%%%%%%%%%%%%%%%%%%%%%%%%%%%%%%%%%%%%%%%%%%%%%%%%%%%%%% 
 
\begin{abstract} 
KIC~3230227 is a short period ($P\approx 7.0$ days) eclipsing binary with a very eccentric orbit ($e=0.6$). From combined analysis of radial velocities and {\it Kepler} light curves, this system is found to be composed of two A-type stars, with masses of $M_1=1.84\pm 0.18M_{\odot}$, $M_2=1.73\pm 0.17M_{\odot}$ and radii of $R_1=2.01\pm 0.09R_{\odot}$, $R_2=1.68\pm 0.08 R_{\odot}$ for the primary and secondary, respectively. In addition to an eclipse, the binary light curve shows a brightening and dimming near periastron, making this a somewhat rare eclipsing heartbeat star system. After removing the binary light curve model, more than ten pulsational frequencies are present in the Fourier spectrum of the residuals, and most of them are integer multiples of the orbital frequency. These pulsations are tidally driven, and both the amplitudes and phases are in agreement with predictions from linear tidal theory for $l=2, m=-2$ prograde modes.
\end{abstract}

%The binary light curve shows a brightening (reflection effect) followed by an eclipse at periastron.

%\keywords{stars:binaries:spectroscopic $-$methods: statistical: bayesian  
%$-$ stars:binaries:eclipsing   
%}
 
%%%%%%%%%%%%%%%%%%%%%%%%%%%%%%%%%%%%%%%%%%%%%%%%%%%%%%%%%%%%%%% 
 
%\setcounter{footnote}{0} 

\section{Introduction}                                % Section 1
Heartbeat stars (HBs), named after the resemblance between their light curves and an electrocardiogram, are binary or multiple systems with very eccentric orbits. The HBs that have been studied in detail include a late B-type star (Maceroni et al.\ 2009), A or F-type stars (Handler et al.\ 2002; Welsh et al.\  2011; Hambleton et al.\ 2013, 2016; Smullen \& Kobulnicky 2015), and red giant stars (Beck et al.\ 2014; Gaulme et al.\ 2013, 2014). Recently, Shporer et al.\ (2016) presented spectroscopic orbits for 19 single-lined HBs. The {\it Kepler} eclipsing binary catalog (Kirk et al.\ 2016) contains over 150 of these stars with the flag `HB'. The distribution of eccentricity ($e$) and orbital period ($P$) for {\it Kepler} eclipsing binaries (EBs) and the 19 HBs is shown by Shporer et al.\ (2016), who note that the HBs occupy the upper envelope of the $(P,e)$ diagram.

The distributions of orbital period and $T_{\rm eff}$ of over 150 HBs in {\it Kepler} EB catalog are shown in Figure 1. The effective temperatures are taken from Armstrong et al.\ (2014). The majority of HBs seem to have orbital period shorter than $30$ days. Their range of effective temperatures ($\sim 5000-7500$ K) suggests that most of them are of spectral type earlier than G (mostly G, F, and A).

KIC~3230227 (HD181850, BD+38 3544; $K_p=9.002$, $\alpha_{2000}$=$19$:$20$:$27.0253$, $\delta_{2000}$=$+38$:$23$:$59.459$) is an eclipsing binary, first included in the {\it Kepler} EB catalog by Slawson et al.\ (2011) and Prsa et al.\ (2011). The original catalog listed the time of eclipse minimum and orbital period as $T_0=54958.702188$ (BJD-2,400,000) and $P=14.094216$ days, respectively. Later, the period was found to be half of the original value ($P=7.0471062$ days).
Uytterhoeven et al.\ (2011) analyzed the {\it Kepler} light curves of $\sim 750$ A- and F-type stars. Among them, KIC~3230227 was classified as an eclipsing binary with $\gamma$ Doradus pulsations. Thompson et al.\ (2012) studied light curves of 17 heartbeat stars, including KIC~3230227. Thanks to the special light curves of HBs, they derived orbital parameters including the orbital inclination ($i$), eccentricity ($e$), and argument of periastron ($\omega_p$). 
Armstrong et al.\ (2014) derived the effective temperatures of $9341\pm 350$K and $7484\pm 606$K for the primary and secondary, respectively, by fitting the SED (Spectral Energy Distribution) to the observed magnitudes. Niemczura et al.\ (2015) made a detailed analysis of their high resolution spectra of KIC~3230227.
Atmospheric parameters were inferred from Na D, H Balmer, and metal lines. They found $T_{\rm eff}=8150\pm 220$K (from the Na D lines and SED), $T_{\rm eff}=8200\pm 100$K (from Balmer and metal lines), $\log g=3.9\pm 0.1$ and $v \sin i=50\pm 4$ km s$^{-1}$. They also obtained abundances for many individual elements (C, N, O, Ne, Na, Mg, etc), as listed in their Table 4. Most of these abundances are close to solar values (Asplund et al.\ 2009; Lodders et al.\ 2009). We summarize the aforementioned results in Table 1.

\section{Binary Modeling}  
The orbital parameters of KIC~3230227 were derived from radial velocity (RV) measurements in Smullen \& Kobulnicky (2015). This system was found to be composed of two A-type stars with similar masses (mass ratio $q = 0.95\pm 0.05$), and a very eccentric orbit ($e=0.60\pm0.04$, $\omega_p=293\pm4 ^{\circ}$). These orbital elements are also listed in Table 1.

Four orbital elements ($P$, $i$, $e$, and $\omega_p$) were derived from the light curve alone by Thompson et al.\ (2012). It is important to note that the Kumar light curve model (Kumar et al.\ 1995) adopted in Thompson et al.'s work does not take into account the effects of reflection and eclipses. The orbital parameters derived by fitting the light curve with the Kumar model can be treated as good estimates, but they can also be off by a large margin, especially the orbital inclination if the reflection effect is important and/or eclipses occur.

A better treatment of the light curve modeling of HBs was performed for the face-on system KOI-54 by Welsh et al.\ (2011). These authors modeled the light curve and radial velocity curve simultaneously, taking advantage of their binary modeling tool ELC (Orosz \& Hauschildt 2000). Stellar distortions were fully modeled with the Roche equipotential, and the reflection effect from mutual heating, plus the limb and gravity darkening effects are included. To synthesize the binary light curve, NextGen atmosphere models are used to integrate numerically  the flux from the stellar surface. Several techniques are adopted in ELC to improve the integration accuracy, for example, Monte Carlo sampling on the fractional pixels at the eclipse horizon with Sobol sequences. Here we use the same tool to model KIC~3230227.

The {\it Kepler} SAP light curves were retrieved from MAST. There are 18 quarters (Q) of long cadence data (Q0-17). Short cadence light curves are only available from quarters 1, 2, and 5. We de-trended the raw light curve in each quarter following the procedure detailed in Guo et al.\ (2016). In short, the procedures include spline fitting to the long term trends, median difference corrections, outlier removal, and normalization. The de-trended light curves were then divided into six temporal sections and light curve modeling was performed for each section individually.

Obvious oscillations stand out in the light curves, and they are still present in the phase-folded light curves (Figure 2, 4). Their amplitudes are low enough to be treated as perturbations to the binary light curve. We adopted the period in the {\it Kepler} Eclipsing Binary Catalog ($P=7.0471062\pm 0.0000175$ days), which is based on the analysis of light curve by using the Lomb-Scargle periodogram and {\it kephem} software (Hambleton et al.\ 2013).

This system only shows a single, very narrow eclipse ($\Delta \phi \approx 0.02$ in phase) near periastron. In order to model fully the shape of the eclipse, we have to use a very small increment in phase ($\delta \phi=0.00055=0\fdg 2$). This makes the light curve computation relatively expensive.
Aperture contamination parameters $k$ listed in Kepler Input Catalog (KIC), which are the percent of contamination light from
other stars in the photometric aperture, range from $0.08\%$ to $0.2\%$. In ELC, this effect is corrected by adding to the median value of the model light curve $y_{\rm med}$ an offset $ky_{\rm med}/(1-k)$. In practice, this effect is usually very small and negligible (Hambleton et al.\ 2016) and we found no significant differences neglecting this effect.
Based on the effective temperatures listed in Table 1, the two components are likely to have radiative envelopes, thus the
gravitational darkening coefficients $\beta_1$ and $\beta_2$ are set to the canonical values of $0.25$ (von Zeipel 1924), and bolometric albedos $l_1,l_2$ are fixed to $1.0$.

%$f_{rot}=\sqrt{\frac{(1+e)}{(1-e)^3}}f_{orb}$ 

The rotational axis of the star is assumed to be aligned with the orbital angular momentum. We assume pseudo-synchronous rotation, which suggests the rotational frequency satisfies $f_{\rm rot}=(1+7.5e^2+5.625e^4+0.3125e^6)/[(1+3e^2+0.375e^4)(1-e^2)^{1.5}]f_{\rm orb}$ (Hut 1981) ($f_{\rm rot}=4.08f_{\rm orb}$ for $e=0.6$). Claret \& Gimenez (1993) showed that early-type binaries exhibit a considerable tendency towards pseudo-synchronism up to $a/R\sim 20$. For KIC3230227, a/R=11.76, and so pseudo-synchronous is a reasonable assumption and it also roughly agrees with the measured $v \sin i$. We proceed here and in Section 3 assuming spin-orbit aligned pseudo-synchronous rotation, but caution that different spin rates/obliquities are possible. We use the orbital eccentricity $e=0.60$ and argument of periastron $\omega_p=293^{\circ}$ from Smullen \& Kobulnicky (2015) as initial values. The mass ratio $q$ and primary semi-amplitude velocity $K_1$, taken from the same paper, are initially fixed to $0.95$ and $98.5$ km s$^{-1}$, respectively. It is well known that the light curves of eclipsing binaries are sensitive to the temperature ratio rather than individual temperatures. Thus the effective temperature of the primary $T_{\rm eff1}$ is fixed to $8000$K, in agreement with the spectroscopic results in Table 1. We fit the light curve by optimizing the following parameters: $e$, $\omega_p$, $i$, relative radius $r_1=R_1/a$ and $r_2=R_2/a$, time of periastron passage $T$, and effective temperature of the secondary $T_{\rm eff2}$. The search for the $\chi^2$ minimum was performed with the genetic algorithm {\it pikaia} (Charbonneau 1995), followed by a local search with the downhill simplex algorithm {\it amoeba}. Since ELC does not model pulsations in the light curve, the standard way of estimating uncertainties by finding the range of parameters that increases $\chi^2$ by 1.0 from $\chi^2_{min}$ cannot be used. Instead, we adopted the standard deviations of the best-fitting parameters in the six data sets as the $1 \sigma$ errors. This is the method used by Guo et al.\ (2016), and it can account for possible systematic uncertainties due to light curve de-trending.

The final optimized solution has essentially the same $e$ and $\omega_p$ values as those in the RV work of Smullen \& Kobulnicky (2015). The orbital inclination ($i=73\fdg 42$), however, is much larger than the result in Thompson et al.\ (2012) ($i=43^{\circ}$), and close to that in Smullen \& Kobulnicky (2015) ($i\sim 66^{\circ}-71^{\circ}$). As shown in Figure 2, our light curve model matches the observations down to the level of $0.001$ magnitude. The profile of the narrow eclipse is also well modeled. The secondary has a slightly higher effective temperature $T_{\rm eff2}/T_{\rm eff1}=1.02$ and smaller mass and radius ($M_2=1.73M_{\odot}, R_2=1.68R_{\odot}$), compared to that of the primary ($T_{\rm eff1}=8000$K, $M_1=1.84M_{\odot}, R_1=2.01R_{\odot}$). The main parameters of our ELC model are listed in Table 2. The projected rotational velocities ($v \sin i_1=56.4$ km s$^{-1}$, $v \sin i_2=47.0$ km s$^{-1}$), under the assumption of pseudo-synchronous rotation, are in approximate agreement with the measured $v \sin i$ from spectra as listed in Table 1. We found that the best-fit light curve solution from one data set or quarter can almost match the light curve of other quarters equally well. In terms of argument of periastron $\omega_p$, no discernable apsidal motion was found.

The flux-weighted radial velocity curves from our ELC model are shown in Figure 3, matching the original RV measurements in Smullen \& Kobulnicky (2015) very well. In the right two panels, we also show the predicted Rossiter-McLaughlin effect  during the eclipse. It can be seen that in order to measure this effect, an RV precision better than $0.5$ km s$^{-1}$ is needed.

\section{Pulsation Characteristics}
\subsection{Tidally Induced Pulsations}

To study the pulsations, we obtained the residuals by subtracting the best binary light curve model from the observations. Figure 4 illustrates the pulsational light curve in Q14, together with the original binary light curve in Q14 and the short cadence light curve in Q5. We then calculated the Fourier spectrum by using the {\it Period}$04$ package (Lenz \& Breger 2005). A standard pre-whitening procedure was performed for the spectrum in each quarter, which means repeating iteratively the following steps: fitting a sinusoid to the data,
subtracting this optimized fit from the data, and computing the Fourier spectrum of the residuals. The fitting formula used is $Z+\sum_i A_i \sin (2\pi (f_i t+\delta_i))$, where $Z, A_i,f_i, \delta_i$ are the zero-point shift, amplitudes of pulsations, frequencies, and phases, respectively. The time $t$ is with respect to the periastron passage: $t= $ BJD $-2,454,958.791621$. The calculation was performed to the long Nyquist frequency ($24.47$ d$^{-1}$). A similar calculation was performed with the short cadence residuals as well, but no peaks were found beyond the frequency $\approx 10$ d$^{-1}$ in the spectrum. The strong pulsational frequencies are actually all below $5$ d$^{-1}$. 

The amplitude spectrum calculated from residuals of quarter 1 only and from quarters $0-17$ are shown in Figure 5. The dominant feature in the spectrum is the equal spacing of the frequency peaks.  The main pulsational frequencies and their amplitudes and phases are listed in Table 3. The uncertainties are calculated following Kallinger et al.\ (2008). We have labeled them as $f_1$ to $f_{10}$, in the order of increasing frequency. A close examination reveals that most of these peaks are exact integer multiples of orbital frequency ($f_{\rm orb}=1/7.0471062= 0.141902 $d$^{-1}$), for instance, $f_3$, $f_5$, $f_6$, $f_7$, $f_8$, $f_9$, and $f_{10}$. The orbital harmonic nature of the pulsational frequencies, together with the high eccentricity of the binary and the masses of the stars strongly suggest that these are tidally induced pulsations. Note that the two non-orbital-harmonic frequencies $f_1$, $f_2$ can be added together to obtain an orbital harmonic ($f_1+f_2=9.88f_{orb}+12.12f_{\rm orb}=22f_{\rm orb}$). The same phenomenon was also found in the tidal oscillations frequencies of KOI-54. This can be explained by non-linear mode coupling as detailed by Burkart et al.\ (2012), O'Leary \& Burkart (2014), and Weinberg et al.\ (2013). It is also interesting to note that $f_1=9.88f_{\rm orb}$ and $f_4=13.88f_{\rm orb}$ have the same fraction to the nearest orbital harmonic. The feature that nonharmonic frequencies share the common fractional parts in units of orbital frequency was discussed in detail by O'Leary \& Burkart (2014) for KOI-54. This further supports the interpretation of these frequencies as the result of non-linear mode coupling. Note that we focus on the significant  frequencies ($S/N >4$) that appear both in the spectra of single-quarter data and all-quarter data. Other tidally induced pulsations with low amplitudes could also exist. For example, the peaks at $0.7095$ d$^{-1}$ and $4.3990$ d$^{-1}$ are $5$ and $31$ times the orbital frequency, respectively.

%and the harmonic peaks generate side-lobes at $Nf_{\rm orb}$

Many frequency triplets can be seen in the spectrum, with equal spacing of orbital frequency. Close examination reveals that all these triplets have frequencies that are equal to $N-0.12, N, N+0.12$ times orbital frequency. Thus they can be explained as a combination of one real oscillation peak and two side-lobes due to the spectral window (Figure 5). The nonharmonic peaks $f_1, f_2, f_4$ generate side-lobes at $(N-0.12)f_{\rm orb}$ and  $(N+0.12)f_{\rm orb}$. We find that after pre-whitening $f_1, f_2$, and $f_4$, these triplets essentially disappear, and only $Nf_{\rm orb}$ peaks remain, supporting the above argument. Low amplitude $m=0$ modes can also exist. At high inclination, these modes are expected to have low amplitudes. As discussed below, the Ledoux constant $C_{nl}$ (Ledoux 1951) is about $0.16$ for the $g$-modes in the observed frequency range. This means the splitting $\delta f$ is about $m(1-0.16) f_{\rm rot}$ for modes with frequencies much higher than $f_{\rm rot}$. If we adopt pseudo-synchronous rotation $f_{\rm rot}=4.1f_{\rm orb}=0.58$ d$^{-1}$, $\delta f$ are then $0.49$ and $0.98$ d$^{-1}$ for $m=1$ and $m=2$, respectively. Thus the splittings will be located at several orbital harmonics away from their central $m=0$ peak. This also suggests that the splittings are not due to rotation.

The amplitude variations of these oscillations are shown in Figure 6 and listed in Table 4. Most of the frequencies have  relatively stable amplitudes over 16 quarters, with variations less than $0.05$ milli-mag. The exception seems to be $f_3$, which decreased from $0.174$ milli-mag in Q1 to $0.078$ milli-mag in Q16.

\subsection{Mode Identification from Phases}

For tidally induced oscillations in KIC 3230227, it is sufficient to consider only the dominant $l=2$ term since higher order terms are at most $(R_1/a)=0.076$ times (an order of magnitude) smaller than the $l=2$ term (eq. 8 in Burkart et al.\ 2012). Additionally, the observed amplitudes of higher order modes ($l\geq 3$) are reduced by geometrical cancelation upon integration over the stellar surface, so we expect their contribution to the pulsation spectrum to be small. 

Phases of tidal oscillations contain important information on the mode properties. O'Leary \& Burkart (2014) identified the two dominant pulsations in KOI-54 as $l=2, m=0$ modes by studying their phases. Following their treatment, for standing modes, the phases of observed flux variation ($\delta J/J$) due to tidal oscillations that have $N$ times the orbital frequency are:

\begin{equation}
\delta=\arg(\delta J_N/J_N)=\arg(A_{nlmN})+\arg(Y_{lm}(\theta_0,\phi_0))=\psi_{nlmN}+m\phi_0
\end{equation}
where, $A_{nlmN}$ is the mode amplitude for the $N$th orbital harmonic (eq. $7$ in Burkart et al.\ 2012) and ($\theta_0,\phi_0)$ are observer's coordinates.

Since we express the pulsations with the formula $\sin (2\pi (f_i t+\delta_i))$ instead of cosine functions, the observed phases ($\delta$) in units of $2\pi$ are then :
\begin{equation}
\delta=\left(\frac{1}{4}+\psi_{nlmN}+m\phi_0 \right)\mod \ \frac{1}{2}
\end{equation}
and 

\begin{equation}
\phi_0=\frac{1}{4}-\frac{\omega_p}{2\pi}
\end{equation}
where $\omega_p=293^{\circ}=5.114$ rad is the argument of periastron from the RV and light curve analysis. In the limit of poor tuning, that is, the difference between intrinsic mode frequency of free oscillations and the nearest orbital harmonics ($\delta\omega=\omega_{nl}-N\Omega_{orb}$) is much larger than mode damping rate ($\gamma_{nl}$), $|\delta \omega| \gg \gamma_{nl}$, we have the following approximation,
\begin{equation}
\psi_{nlmN} \approx \left[ \pi/2-\arctan \left(\frac{\delta \omega}{\gamma_{nl}}\right) \right]/(2\pi)=\left[\pi/2-\pi/2\right]/(2\pi)=0
\end{equation}
The observed phase is then 
\begin{equation}
\delta=\left( \frac{1}{4}+m\phi_0 \right) \ \   \mod \ \frac{1}{2}
\end{equation}

%As the time dependence of these orbital harmonic tidal oscillations is $\sin[2\pi(|N|f_{\rm orb}t+\delta]$ and we can not determine the sign of $N$ (we can only observe $|N|$), we refrain the specification of these modes as prograde or retrograde (Burkart et al.\ 2012). 
%We conclude that the observed oscillations are very likely due to $l=2,m=|2|$ modes

Note that if using the magnitude variation, the phase will be off by $\pi$ (by $1/2$ if in units of $2\pi$), since $\delta {\rm mag}\propto -\delta J/J$. In Figure 7, we show the observed phases of main oscillations. Within uncertainties, phases of $f_2, f_3, f_5, f_6, f_7, f_8, f_9, f_{10}$ can be explained by the theoretical phases of $l=2, m=-2$ modes ($\delta_{m=-2}=0.38,     0.88$). The phase of $f_4$ is close to the predicted phase of $m=2$ modes ($\delta_{m=2}=0.12$). 
Thus most of the observed oscillations 
are likely due to $l=2,m=-2$ 
modes, in agreement with the expectations for the high inclination angle of the binary ($i=73\fdg 4$). On the other hand, $l=2, m=0$ modes are expected to have low amplitudes, and none of the main frequencies have phases close to those predicted phases ($\delta_{m=0}=0.25,0.75$). We do not have strong evidence that the oscillations are in resonance locking, because amplitudes of resonance-locked modes are much larger than normal tidal oscillations (see the amplitude modeling below).

%The observed phases thus seem to rule out this possibility.

\subsection{Theoretical Flux Variations}

We want to study whether the observed amplitudes of tidal oscillations agree with theory.
To this end, we evolve a star of $M=1.84M_{\odot}$ with solar metallicity with MESA evolution code (Paxton et al.\ 2011, 2013) until its properties match the observations of the primary. The closest equilibrium model has the same radius ($R=2.01R_{\odot}$), but slightly cooler effective temperature ($T_{\rm eff}({\rm model})=7800$K vs.  $T_{\rm eff}({\rm observation})=8000$K). We use parameters of the secondary ($M_2=1.73M_{\odot}$) for the calculation of tidal forcing from the companion. 

Following Fuller \& Lai (2012), the Lagrangian tidal displacement $\bm{\xi}(\bm{r},t)$ can be expressed as the sum of displacement of free oscillations $\bm{\xi}_{\alpha}(\bm{r})$, 

\begin{equation}
\begin{split}
&\bm{\xi}(\bm{r},t)=\sum_{\alpha} c_{\alpha}(t) \bm{\xi}_{\alpha}(\bm{r}) \\
\end{split}
\end{equation}
where $\alpha$ represents the mode indices which include ($n, l, m$). The amplitude of each mode $c_{\alpha}$ is derived from solving the forced harmonic oscillator equation (their eq. 22), and the solution is given by their eq. 23. The expression of $c_{\alpha}(t)$ involves the sum over the forcing from each orbital harmonic $N$, and this is from the Fourier expansion of orbital motion in the eccentric orbit (their eq. 20). The displacement $\bm{\xi}_{\alpha}$ and various other eigenfunctions of $l=2, m=0$ modes are calculated with the GYRE code (Townsend \& Teitler 2013) in the non-adiabatic mode.
We use the updated collocation method COLLOC\_GL2 to solve the oscillation equations which has better performance than the Magnus solver for non-adiabatic calculations. 

We use the perturbative approximation which is valid when (angular) rotation frequency $\Omega_{\rm rot}$ is much smaller than mode frequency (co-rotating frame) in the zero-rotation limit ($\omega_{nl}$). The frequencies of $l=2, m=-2$ prograde modes are calculated from $\omega_{\alpha}=\omega_{nlm}=\omega_{nl}+mC_{nl}\Omega_{\rm rot}$. The mode eigenfunctions are normalized to have unit mode inertia and assumed to be unchanged by rotation.

Following Buta \& Smith (1979), the magnitude variation of a single oscillation mode due to temperature changes has the following expression assuming pulsations are adiabatic:
\begin{equation}
\begin{split}
&(\Delta {\rm mag})_{T}\\
&=-1.0857\frac{xe^x}{e^x-1}\left[ \frac{\xi_r(R)}{R} \frac{\Gamma_2-1}{\Gamma_2}\left(\frac{l(l+1)}{\omega^2}-4-\omega^2\right)\right]\gamma_{l} \ \sqrt{\frac{(2l+1)(l-m)!}{4\pi(l+m)!}}P_{l}^m(\cos i_s)e^{im\phi_0}e^{i\omega_{\alpha} t}\\
%&=-1.0857\frac{xe^x}{e^x-1}\left[ \frac{\delta T}{T}\right]\gamma_{l} \ \sqrt{\frac{(2l+1)(l-m)!}{4\pi(l+m)!}}P_{l}^m(\cos i_s)e^{im\phi_0}e^{-i\omega_{\alpha} t}\\
\end{split}
\end{equation}
where  $\xi_r(R)$ is the radial displacement evaluated at the stellar surface. The term in the square bracket $[\ ]$ is the approximation to the temperature perturbation at the stellar surface $\frac{\delta T}{T}|_R$, and $\frac{xe^x}{e^x-1}$ arises from the blackbody approximation to the stellar atmosphere, with $x=hc/\lambda kT$. $i_s$ is the orbital inclination, $\Gamma_2 \approx 5/3$ is the adiabatic index, and $\omega$ is the dimensionless mode frequency given by $\omega=\omega_{\alpha}/\sqrt{GM/R^3}$.
$\gamma_l$ is bolometric limb darkening coefficient defined in eq. (39) of Buta \& Smith (1979). For an A-star similar to KIC~3230227, $\gamma_l$ is about $0.3$ in the {\it Kepler} passband. The above equation is good for the first-order approximation of magnitude variations, and a better treatment should fully take into account the non-adiabaticity of oscillations (J. Fuller 2016, in prep.). The variations due to geometric changes are usually much smaller and thus are not considered here.

%\textbf{The theoretical amplitudes show a peak at the harmonic $N=8$, and this is due to the numerical disability of equation (7) since the mode frequencies $\omega$ approach zero and the square bracket term goes to infinity.}

Using eq.\ (7) and summing up the contribution from each mode $\alpha$, we calculated the magnitude variation for each orbital harmonic $N$ for $l=2, m=-2$ prograde modes. The variations due to the equilibrium tide (setting $N\Omega_{\rm orb}=0$) have been subtracted. The result is shown in Figure 8, together with the observed amplitudes of oscillations.  The predicted mode amplitudes are very sensitive to the mode detuning parameter $\delta \omega=\omega_{\alpha}-N\Omega_{\rm orb}$, i.e., the difference between the intrinsic mode frequency $\omega_{\alpha}$ and driving frequency $N\Omega_{\rm orb}$. There is a strong peak at $N=23$, which is due to a chance resonance (very small $\delta \omega$) for the stellar model that we use. A stellar equilibrium model with almost the same observed parameters (radius, temperature, and mass) and slightly different mode frequencies will have quite different mode detuning. Detailed amplitude modeling requires very fine grids of structure models and is beyond the scope of this paper (see Burkart et al.\ 2012). But overall, the theoretical predicted mode amplitudes seem to agree with observations. It further supports the argument that the oscillations are due to tidally excited $m=-2$ quadruple modes.

\section{Summary and Future Prospects}
The unprecedented light curves from the {\it Kepler} satellite offer us opportunities to study the effect of tides on stellar oscillations. Heartbeat stars in eclipsing systems are 
among the best laboratories since the model independent fundamental
stellar parameters such as mass and radius can be determined. We presented a study of KIC~3230227, which consists of two
A-type stars in an eccentric orbit with a period of 7 days. The observed
pulsations, mostly orbital harmonics, can be explained by the tidally induced $l=2, m=-2$ prograde modes. This is supported by a comparison of their observed and modeled phases and amplitudes. 

The fundamental parameters of KIC~3230227 are determined only to $10\%$ in mass and $5\%$ in radius. Further analysis could take advantage of the high resolving power spectra and more phase coverage in the RV curve. This is already underway \footnote{While our paper was under review, we learned about the preliminary work by Lampens et al.\ presented as a poster at the KASC8/TASC1 meeting. The orbital parameters derived from their high resolving-power spectra agree with our result. They found that this binary is actually a triple system, and the estimated flux contribution of the third star at $5300$\ \AA \ is about $4\%$. Our light curve solution may change slightly with the inclusion of a small third light, but it will not change our conclusions about the pulsational properties. However, the third star could have a great influence on the evolution history of orbital parameters (eccentricity, spin-oribit alignment, and rotation rate) for this system.} (K. Hambleton, private communication). Once more accurate parameters are determined, asteroseismic modeling of these tidal oscillations can be performed, as was done in Burkart et al.\ (2012). To solidify the result of this work, mode identification techniques can be applied to the line profiles variations as well as to the time series of multi-color photometry.
It is also worthwhile studying the Fourier spectrum more closely, identifying individual modes, and analyzing the non-linear mode couplings. Another weakness of this work is that we are unable to tell which star is pulsating or if both stars are pulsating. A study of pulsations during eclipse may help to clarify this issue (B{\'{\i}}r{\'o} \& Nuspl 2011).

%%%%%%%%%%%%%%%%%%%%%%%%%%%%%%%%%%%%%%%%%%%%%%%%%%%%%%%%%%%%%%% 
 
\acknowledgments 
 
%We thank the anonymous referees for helpful comments and suggestions which greatly improve the quality of this paper. 
We thank the anonymous referee for helpful comments and suggestions which  improve the quality of this paper. We thank Jerome A. Orosz for making his ELC code available to us. We thank Bill Paxton, Rich Townsend and others for maintaining and updating MESA and GYRE. G. Z. is grateful to Joshua Burkart for explaining tidal asteroseismology, to Rachel Smullen for her help in clarifying some issues in the analysis of radial velocities. We thank Kelly Hambleton for useful discussions.
This work is partly based on data from the {\it Kepler} mission. {\it Kepler} was competitively selected as the tenth Discovery mission. Funding for this mission is provided by NASA's Science Mission Directorate.
The photometric data were obtained from the Mikulski Archive for Space Telescopes (MAST). STScI is operated by the Association of Universities for Research in Astronomy, Inc., under NASA contract NAS5-26555. This study was supported by NASA grants NNX12AC81G, NNX13AC21G, and NNX13AC20G. This material is based upon work supported by the National Science Foundation under Grant No. ~AST-1411654. Institutional support has been provided from the GSU College 
of Arts and Sciences and the Research Program Enhancement 
fund of the Board of Regents of the University System of Georgia, 
administered through the GSU Office of the Vice President 
for Research and Economic Development.

%{\it Facilities:} \facility{Kepler, Mayall} 
 
%%%%%%%%%%%%%%%%%%%%%%%%%%%%%%%%%%%%%%%%%%%%%%%%%%%%%%%%%%%%%%% 
% References 

%%%%%%%%%%%%%%%%%%%%%%%%%%%%%%%%%%%%%%%%%%%%%%%%%%%%%%%%%%%%%%% 

\clearpage

%%%%%%%%%%%%%%%%%%%%%%%%%%%%%%%%%%%%%%%%%%%%%%%%%%%%%%%%%%%%%%% 
%from ELC.parm/

% Table 3 
\begin{deluxetable}{lcccccc} 
\rotate
\tabletypesize{\footnotesize} 
\tablewidth{0pc} 
\tablenum{1} 
\tablecaption{Atmospheric and Orbital Parameters\label{tab1}} 
\tablehead{ 
\colhead{Parameter}   & 
\colhead{Uytterhoeven}      &
\colhead{Thompson }      &
\colhead{Armstrong} & 
\colhead{Smullen \&}      &
\colhead{Niemczura} & 
\\
\colhead{}   & 
\colhead{et al.\ (2011)}      &
\colhead{et al.\ (2012) }      &
\colhead{et al.\ (2014)} & 
\colhead{ Kobulnicky (2015) }  & 
\colhead{et al.\ (2015)} & 
}
\startdata 
$T_{\rm eff}$ (K)               \dotfill & $7970(290)$ & $8750 $ & $9341(350)\tablenotemark{a}$,$7484(606)\tablenotemark{a}$ & $\sim 8000$ &$8150(220),8200(100)$\\ 
$\log g$ (cgs)  \dotfill & $3.9\pm 0.3$  & $5.0$  & $-$  & $4.0,3.5$ &3.9(0.1)     \\ 
$v \sin i$ (km s$^{-1}$)     \dotfill & $-$        & $-$   & $-$        & $\sim 30,\sim 75$ &$50(4)$            \\ 
$[{\rm Fe/H}]$ \dotfill & $-$        & $-$  & $-$    & $ \geq 0$  &$\approx 0$           \\ 
\hline
$P$ (days) & $7.0471062(175)\tablenotemark{b}$        & $7.04711(87)$   & $-$        & $7.051(1)$  \\
$T_0$ (BJD-2400000) & $54958.702238\tablenotemark{c}$        & $-$   & $-$        & $-$  \\
$T$ (BJD-2400000) & $-$        & $-$   & $-$        & $56311.76(03)\tablenotemark{d}$  \\
$i$ ($^\circ$) & $-$        & $42.79\pm 0.46$   & $-$        & $66-71$ &$-$ \\
$e$  & $-$        & $ 0.588(4)$   & $-$        & $0.60(4)$  \\
$\omega_p$ ($^{\circ}$) & $-$        & $292.1(1.2)$   & $-$        & $293(4)$  \\
\hline
$K_1$ (km s$^{-1}$)  & $-$        & $ -$   & $-$        & $98.5(5.4)$ &$-$ \\
$K_2$ (km s$^{-1}$)  & $-$        & $ -$   & $-$        & $104.9(6.1)$ &$-$ \\
$\gamma$  (km s$^{-1}$)  & $-$        & $-$   & $-$        & $-15.7(1.7)$ &$-$ \\
\enddata 
\tablenotetext{a}{For the primary and secondary, respectively}
\tablenotetext{b}{Kepler Eclipsing Binary Catalog}
\tablenotetext{c}{Time of eclipse minimum from Kepler Eclipsing Binary Catalog}
\tablenotetext{d}{Time of periastron passage}
\end{deluxetable} 

\clearpage

\begin{deluxetable}{lccc} 
\tabletypesize{\small} 
\tablewidth{0pc} 
\tablenum{2} 
\tablecaption{Model Parameters\label{tab2}} 
\tablehead{ 
\colhead{Parameter}   & 
\colhead{Primary}      &
 \colhead{Secondary}  &
\colhead{System}       
}
\startdata 
Period (days) & & &$7.047106 \tablenotemark{a}\pm 0.000018$\\
Time of periastron passage, $T$ (BJD-2400000) & & &$54958.791621 \pm 0.000010$   \\
Mass ratio $q=M_2/M_1\tablenotemark{b}$               & & &$0.939 \pm 0.075$    \\ 
Orbital eccentricity, $e$              & & &$0.600\pm 0.005$     \\
Argument of periastron, $\omega_p$ (degree)              & & &$293.0\pm 1.0$     \\
$\gamma $ velocity\tablenotemark{b} (km s$^{-1}$)& & &$-15.7\pm 1.7$\\
Orbital inclination (degree), $i$ & & & $73.42\pm 0.27$\\
Semi-major axis ($R_\odot$), $a$ & & & $23.64\pm 0.95$\\
Mass ($M_\odot$)              & $1.84 \pm 0.18$             & $1.73 \pm 0.17$    \\ 
Radius ($R_\odot$)               & $2.01\pm 0.09$       & $1.68 \pm 0.08$    \\

Relative radius, $R/a$     &$0.085\pm 0.002$    &$0.071\pm 0.002$\\
Gravity brightening, $\beta$     &$0.25\tablenotemark{a}$    &$0.25\tablenotemark{a}$\\

Bolometric albedo & $1.0\tablenotemark{a}$   & $1.0\tablenotemark{a}$    \\

$T_{\rm eff}$ (K)                    & $8000\tablenotemark{a} $ & $8177 \pm 30$\\
$\log g$ (cgs)  & $4.10\pm 0.06$     & $4.23\pm 0.06$     \\ 
pseudo-synchronous $v \sin i$ (km s$^{-1}$)   &   $56.4 \pm 1.4$   &   $47.0 \pm 1.1$              \\ 
Velocity semiamplitude $K$\tablenotemark{b} (km~s$^{-1})$  &$98.5\pm 5.4 $ & $104.9\pm 6.1$\\ 
%rms of $V_r$ residuals (km~s$^{-1})$  &$6.2$ & $9.7$\\ 
\enddata 
\tablenotetext{a}{Fixed.}
\tablenotetext{b}{Adopted from Smullen \& Kobulnicky (2015). }
\end{deluxetable}

%findsn_lcall.pro
% Table 3 
\begin{deluxetable}{lcccccc} 
\tabletypesize{\small} 
\tablewidth{0pc} 
\tablenum{3} 
\tablecaption{Main Oscillation Frequencies of KIC~3230227
\label{tab3}} 
\tablehead{ 
\colhead{}   & 
\colhead{Frequency (d$^{-1}$)}   & 
\colhead{Amplitude ($10^{-3}$mag)}      &
 \colhead{Phase ($2\pi$)}  &
 \colhead{S/N}  &
 \colhead{Comment}  &
}
\startdata 
$f_1$ &$1.40214\pm 0.00002$   & $0.179\pm 0.027 $ & $0.97\pm 0.07$&11.01&$9.88f_{orb}$\\
$f_2$   & $1.71988\pm 0.00002 $ &$0.192  \pm 0.022$  &$0.34 \pm 0.05 $&14.92&$12.12f_{orb}$\\
$f_3$       & $1.84482 \pm 0.00002$ &$0.096\pm   0.021$  &$0.32  \pm 0.10$&7.95&$13f_{orb}$\\
$f_4$       & $1.969765\pm 0.000008 $ & $0.338    \pm 0.020 $& $0.16 \pm  0.03$&29.33&$13.88f_{orb}$\\
$f_5$       & $2.12855\pm 0.00001 $ &$0.188\pm 0.018$& $0.89  \pm 0.05$&17.47&$15f_{orb}$\\
$f_6$       & $2.41235\pm 0.00001 $ & $0.189  \pm  0.016$&$0.39 \pm 0.04$&20.28&$17f_{orb}$\\
$f_7$       & $2.55425\pm 0.00002 $ &$0.118  \pm  0.015$ &$0.85  \pm 0.06$&13.55&$18f_{orb}$\\
$f_8$       & $2.69615\pm 0.00002 $ &$0.159  \pm  0.014$ &$0.37  \pm 0.04$&19.84&$19f_{orb}$\\
$f_9$       & $2.83805\pm 0.00002 $ &$0.076  \pm  0.013$ &$0.37  \pm 0.08$&10.18&$20f_{orb}$\\
$f_{10}$       & $2.979948\pm 0.000008 $ &$0.192  \pm  0.012$ &$0.86  \pm 0.03$&27.52&$21f_{orb}$\\
\enddata 
\end{deluxetable} 

\clearpage

\begin{deluxetable}{lccccccccccc} 
\tabletypesize{\small} 
\tablewidth{0pc} 
\tablenum{4} 
\tablecaption{Amplitude Variations of the Main Frequencies ($10^{-3}$ mag) \label{tab3}} 
\tablehead{ 
\colhead{Quarter}   & 
\colhead{$f_1$}   & 
\colhead{$f_2$}      &
 \colhead{$f_3$}  &
 \colhead{$f_4$}  &
 \colhead{$f_5$}  &
 \colhead{$f_6$}  &
 \colhead{$f_7$}  &
 \colhead{$f_8$}  &
 \colhead{$f_9$}  &
 \colhead{$f_{10}$}  &
}
\startdata 
Q1 &      0.178&      0.177&      0.174&      0.294&      0.201&      0.200&      0.116&      0.163&
     0.074&      0.212\\
Q2 &      0.177&      0.185&      0.126&      0.336&      0.232&      0.192&      0.159&      0.161&
     0.085&      0.215\\
Q3 &      0.177&      0.173&      0.157&      0.320&      0.215&      0.190&      0.155&      0.159
&     0.082&      0.213\\
Q4 &      0.163&      0.202&      0.122&      0.330&      0.215&      0.191&      0.132&      0.165
&     0.074&      0.206\\
Q5 &      0.171&      0.175&      0.119&      0.332&      0.231&      0.171&      0.120&      0.166
&     0.076&      0.208\\
Q6 &      0.178&      0.182&      0.122&      0.319&      0.202&      0.1881&      0.126&      0.164
&     0.075&      0.209\\
Q7 &      0.168&      0.190&     0.094&      0.334&      0.232&      0.182&      0.133&      0.164
&     0.077&      0.205\\
Q8 &      0.172&      0.194&      0.113&      0.319&      0.201&      0.173&      0.121&      0.146
&     0.076&      0.205\\
Q9 &      0.164&      0.189&     0.054&      0.331&      0.218&      0.191&      0.145&      0.157
&     0.081&      0.211\\
Q10 &      0.168&      0.176&     0.073&      0.335&      0.198&      0.193&      0.126&
      0.160&     0.070&      0.205\\
Q11 &      0.160&      0.190&     0.056&      0.333&      0.207&      0.190&      0.129&
      0.164&     0.071&      0.213\\
Q12 &      0.154&      0.203&     0.069&      0.328&      0.227&      0.174&      0.145&
      0.155&     0.064&      0.205\\
Q13 &      0.157&      0.198&     0.061&      0.346&      0.211&      0.191&      0.134&
      0.164&     0.081&      0.208\\
Q14 &      0.178&      0.205&     0.053&      0.331&      0.207&      0.183&      0.134&
      0.158&     0.080&      0.203\\
Q15 &      0.165&      0.190&     0.091&      0.321&      0.193&      0.190&      0.137&
      0.157&     0.080&      0.209\\
Q16 &      0.151&      0.205&     0.078&      0.321&      0.198&      0.184&      0.118&
      0.152&     0.075&      0.199\\
%Q17 &      0.181&      0.155&      0.107&      0.279&      0.168&      0.220&      0.151&
%     0.155&     0.085&      0.222\\
\hline
$1\sigma$ error &0.027 & 0.022 & 0.021 &0.020 &0.018 &0.015 &0.015 &0.014 &0.013 &0.012 \\
\enddata 
\end{deluxetable} 

\clearpage

% Figures 

\begin{figure} 
\begin{center} 
{\includegraphics[height=11cm]{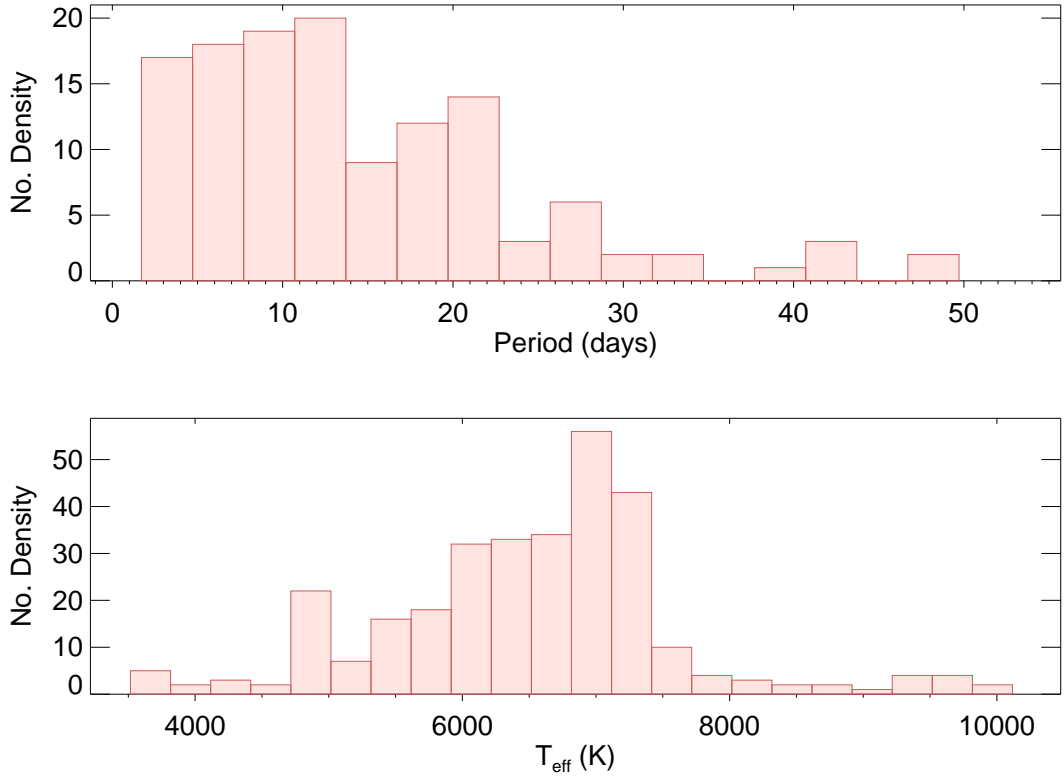}} 
\end{center} 
\caption{Number distribution of orbital period (upper panel) and effective temperature (lower panel) for 157 HBs in {\it Kepler} eclipsing binary catalog.}
\end{figure} 
 
% ms323/checkfit_plot.pro
% Figure 2  
\begin{figure} 
\begin{center} 
{\includegraphics[angle=0,height=14cm]{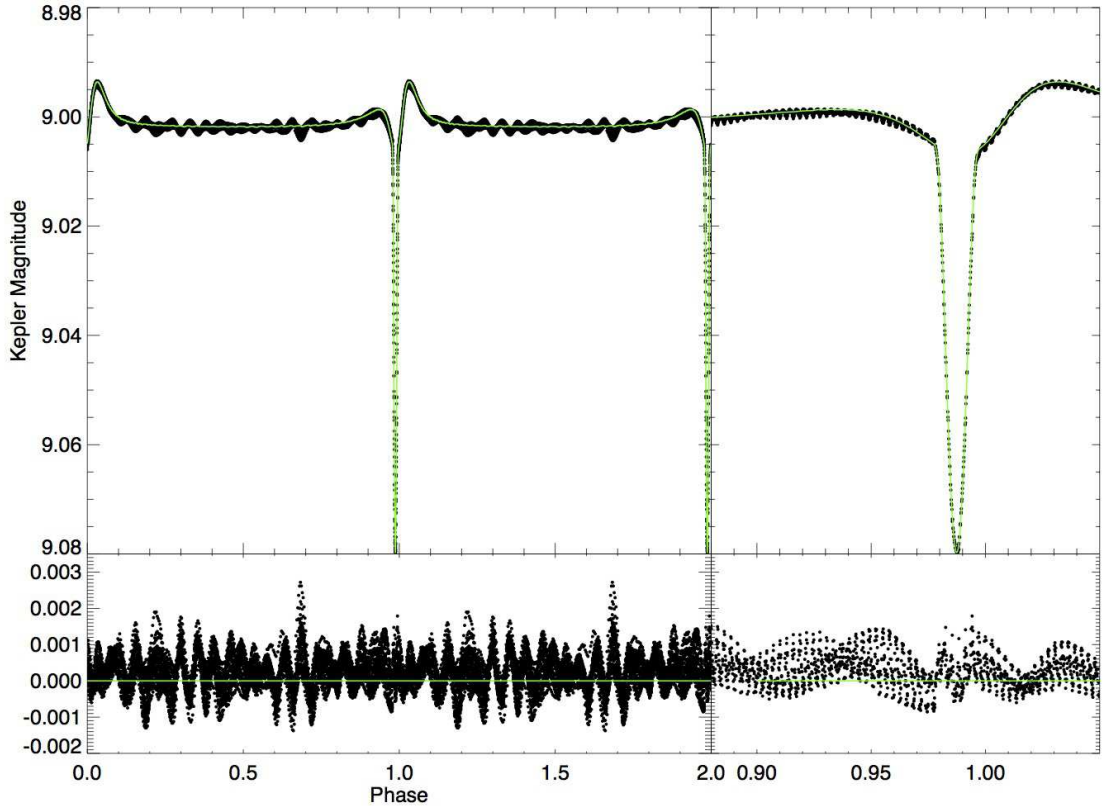}} 
%{\includegraphics[height=12cm]{ew2_diff.eps}} 
\end{center} 
\caption{The phase-folded long cadence light curve of KIC~3230227 (dots) in Quarter 5 and 6 and the best model from ELC (green solid line). The right panel shows the light curve around the eclipses, and the bottom panels show the corresponding residuals. Note that the seemingly high-frequency `oscillations' in the upper-right panel are artifacts of folding the long cadence data and are not real. This is confirmed by the absence of high-frequency peaks in the Fourier spectrum of short cadence light curves. }
\end{figure} 
%the 'high freq' in the upper right panel is artifact from folding,
%plot323sc.pro show sc data, and 'hight freq osc' donot exist

% Figure 3 
\begin{figure} 
\begin{center} 
{\includegraphics[angle=0,height=12.2cm]{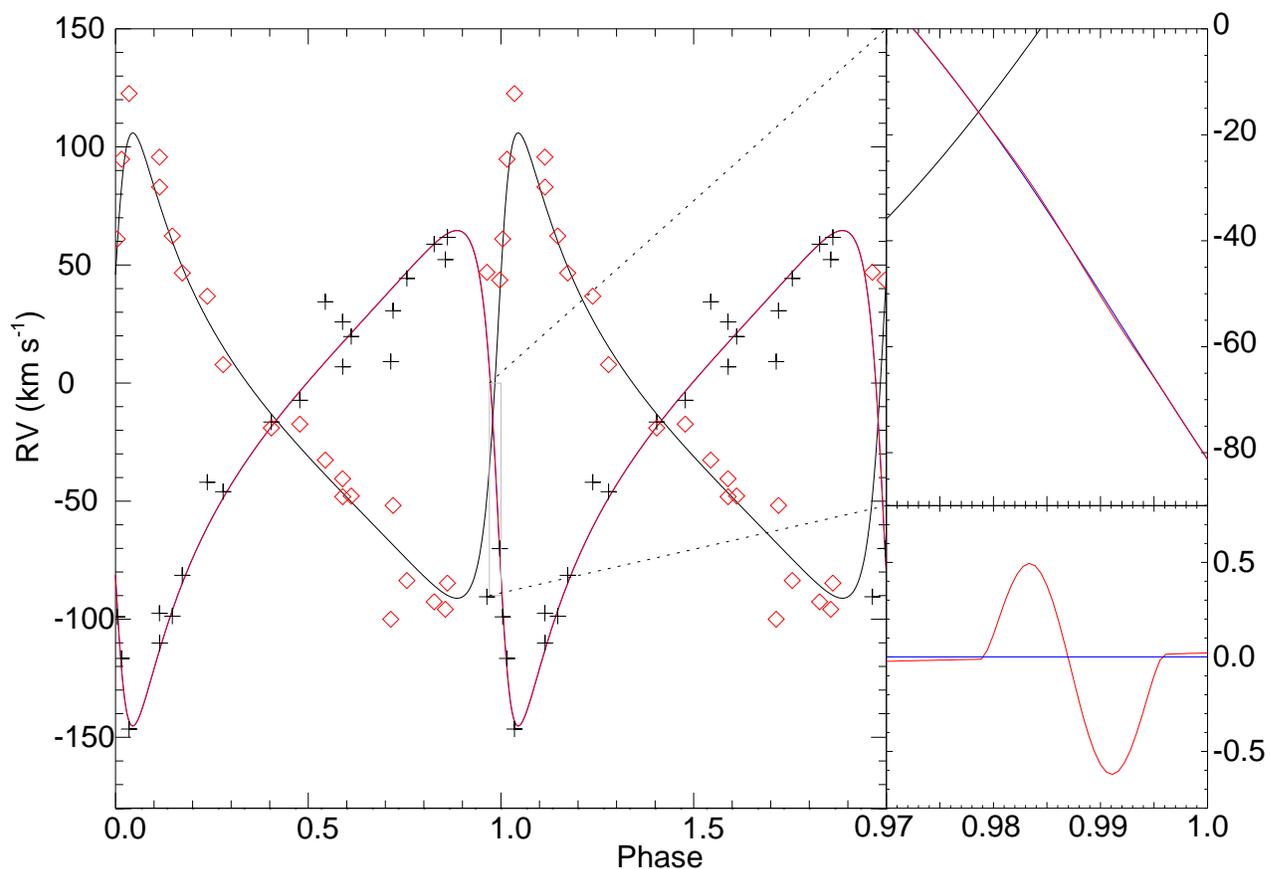}} 
%{\includegraphics[height=12cm]{ew2_diff.eps}} 
\end{center} 
\caption{The radial velocity models of the primary (black solid) and the secondary (red solid) star from ELC. The periastron passage corresponds to phase zero. The corresponding observed radial velocities are indicated as red diamonds and black crosses. The upper right panels shows the RVs during the eclipse. The red curve represents a flux-weighted radial velocity model and blue curve is a simple Keplerian model. The RV residuals of the two models are shown in the lower right panel. }
\end{figure} 

% Figure 3 
\begin{figure} 
\begin{center} 
{\includegraphics[angle=0,height=13.5cm]{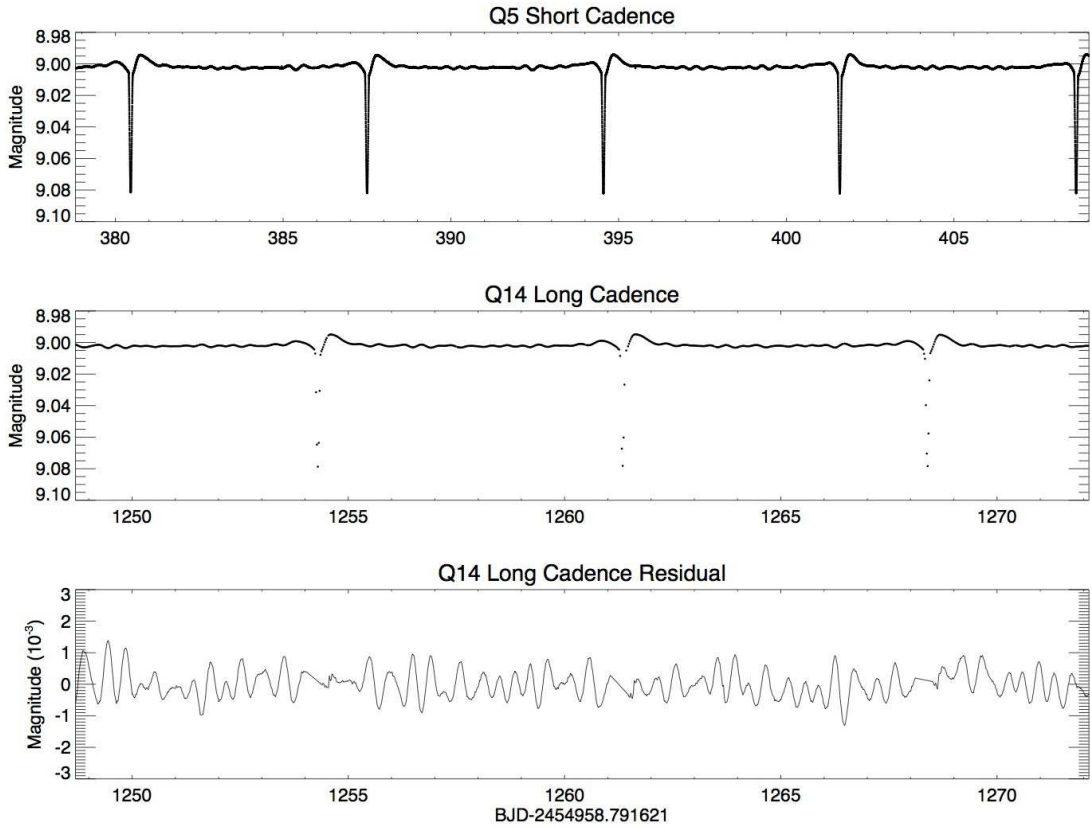}} 
%{\includegraphics[height=12cm]{ew2_diff.eps}} 
\end{center} 
\caption{ \textbf{Upper panel}: The short cadence light curve in quarter 5.\textbf{ Middle panel}: The long cadence light curve in quarter 14. \textbf{Lower panel}: The light curve residuals after subtracting the best-fit binary light curve model in quarter 14. Eclipses have been masked.}
\end{figure} 

% Figure 4 
\begin{figure} 
\begin{center} 
{\includegraphics[angle=90,height=12cm]{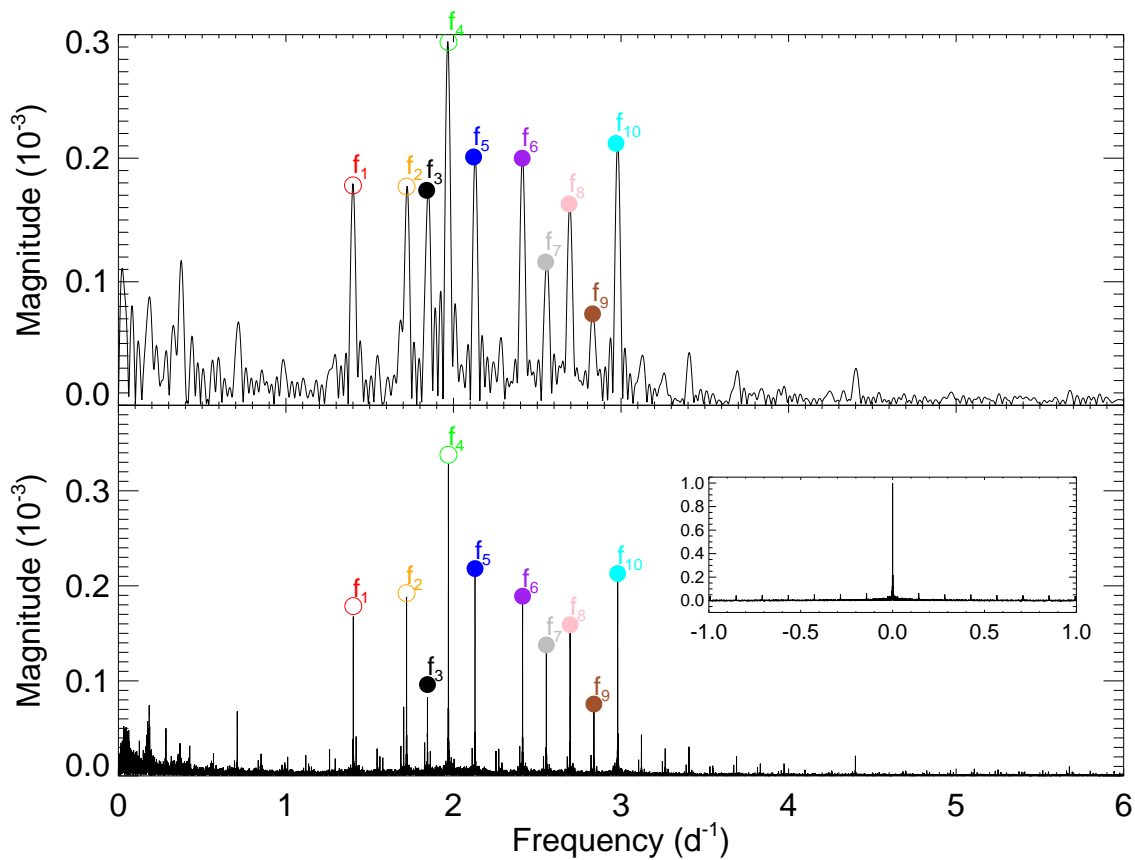}} 
%{\includegraphics[height=12cm]{ew2_diff.eps}} 
\end{center} 
\caption{ Fourier spectrum of light curve residuals with eclipses masked. The upper panel was calculated from the quarter 1 long cadence data. The lower panel presents a similar plot but using all quarters (Q$0-17$) of long cadence data. The 10 dominant frequencies listed in Table 2 are labeled. Filled and open circles mark the harmonic and nonharmonic orbital frequencies, respectively. The spectral window is shown in the inset.}
\end{figure}

% Figure 4 
\begin{figure} 
\begin{center} 
{\includegraphics[angle=90,height=12cm]{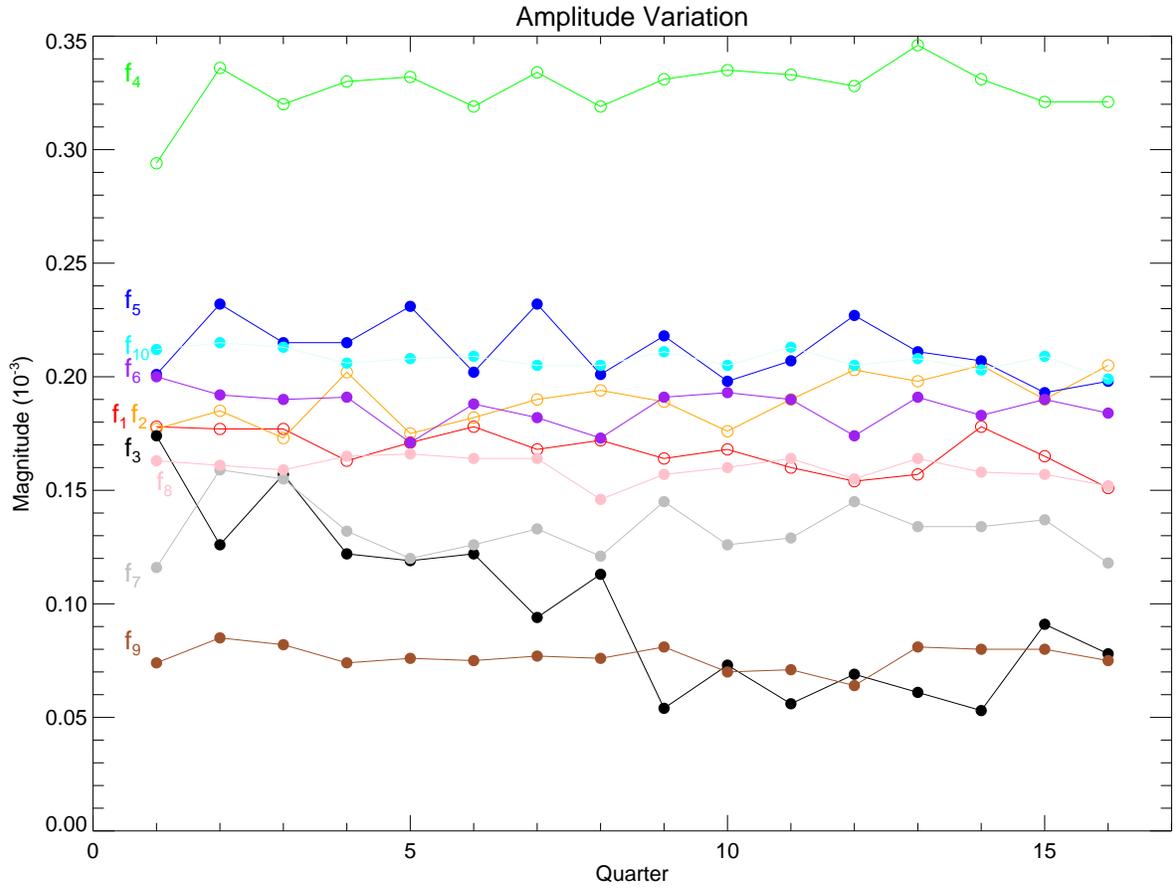}} 
%{\includegraphics[height=12cm]{ew2_diff.eps}} 
\end{center} 
\caption{ The amplitude variations of ten dominant oscillation frequencies (Table 3 and 4). Filled and open circles indicate the harmonic and nonharmonic orbital frequencies, respectively. }
\end{figure}

% Figure 4 
\begin{figure} 
\begin{center} 
{\includegraphics[angle=90,height=12.5cm]{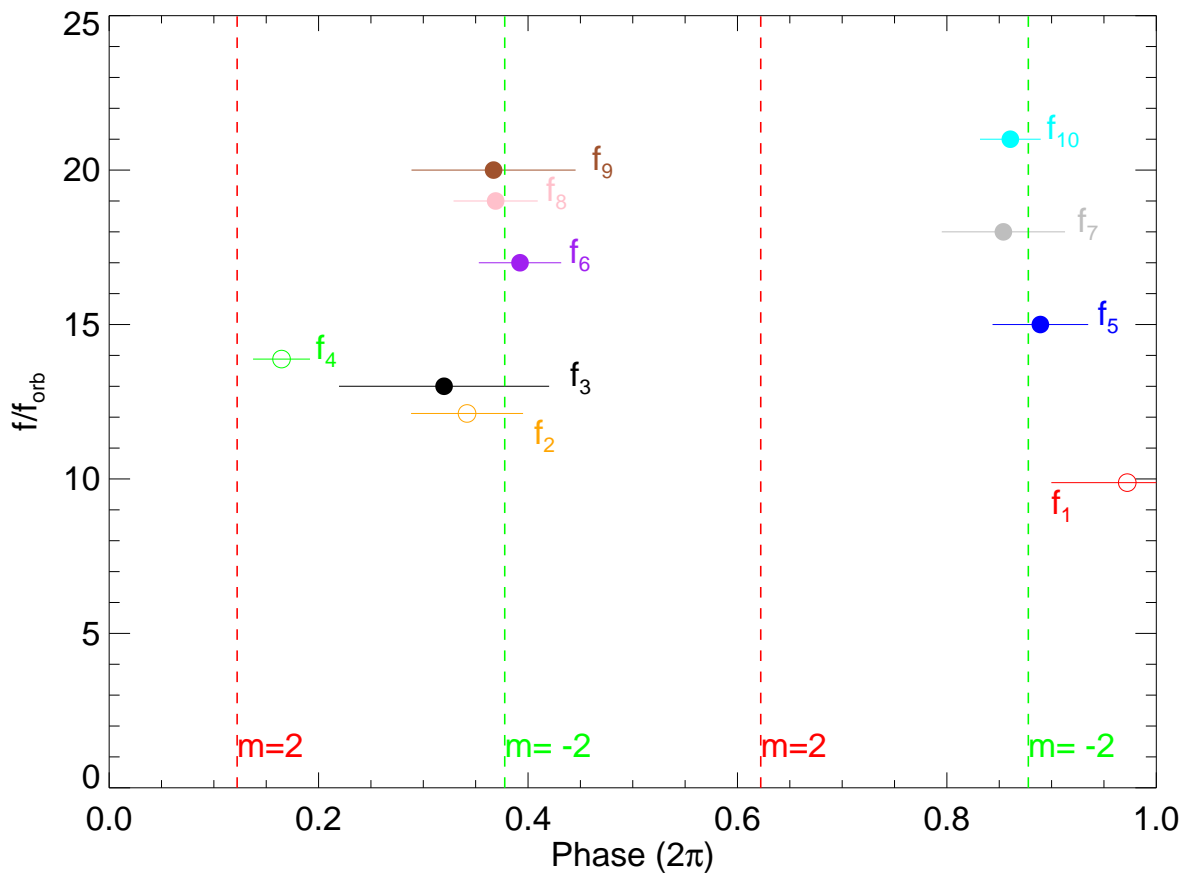}} 
%{\includegraphics[height=12cm]{ew2_diff.eps}} 
\end{center} 
\caption{ Phases of ten dominant oscillations (see Table 3). The $1\sigma$ error bars of phases are shown, and those of frequencies are smaller than the symbols. Red and green dotted lines indicate the theoretical phases of $l=2,m=2$ and $l=2,m=-2$ modes. Filled and open circles indicate the harmonic and nonharmonic orbital frequencies, respectively.}
\end{figure}

%/Users/feynman/Desktop/Numinteg/k323_gyre43_inertia_1/
%k323_m2.pro
% Figure 5 
\begin{figure} 
\begin{center} 
{\includegraphics[angle=0,height=13cm]{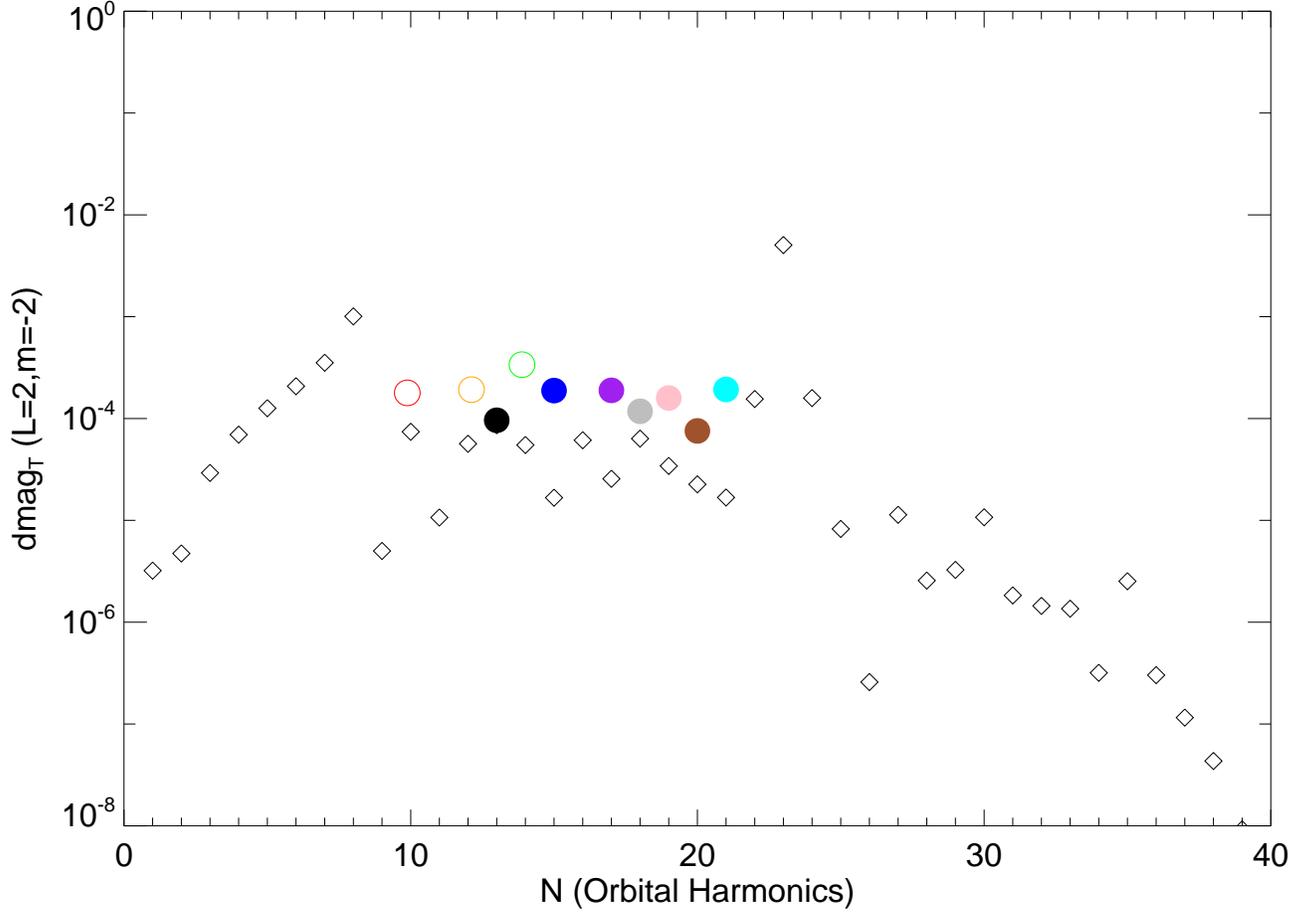}} 
%{\includegraphics[height=12cm]{ew2_diff.eps}} 
\end{center} 
\caption{ Theoretical magnitude variations of $l=2, m=-2$ prograde modes are indicated by diamonds. The observed magnitudes of oscillations are shown as color symbols. Oscillation frequencies that are orbital harmonics are indicated by the filled circles, and otherwise by open circles.}
\end{figure}

\end{document}